\documentclass[conference]{IEEEtran}
\IEEEoverridecommandlockouts
\usepackage{cite}
\usepackage{amsmath,amssymb,amsfonts}
\usepackage{algorithmic}
\usepackage{graphicx}
\usepackage{textcomp}
\usepackage{booktabs} % For formal tables
\usepackage{graphicx} 
\usepackage[draft]{hyperref}
\usepackage{amssymb}
\usepackage[font=footnotesize,labelfont=bf]{caption}
\usepackage[font=footnotesize,labelfont=bf]{subcaption}
\def\BibTeX{{\rm B\kern-.05em{\sc i\kern-.025em b}\kern-.08em
    T\kern-.1667em\lower.7ex\hbox{E}\kern-.125emX}}
\begin{document}

\title{Impact of Semantic Granularity on Geographic Information Search Support
\thanks{This work was funded by the Universities of Turin and Genova, and by University College Dublin.}}

\author{\IEEEauthorblockN{N. Mauro, L. Ardissono}
\IEEEauthorblockA{\textit{Computer Science Department} \\ \textit{University of Torino} \\
Torino, Italy \\
noemi.mauro@unito.it, \\liliana.ardissono@unito.it}
\and
\IEEEauthorblockN{L. Di Rocco, G. Guerrini}
\IEEEauthorblockA{\textit{DIBRIS} \\ \textit{ University of Genova} \\
Genova, Italy\\
laura.dirocco@dibris.unige.it, \\giovanna.guerrini@unige.it}
\and
\IEEEauthorblockN{M. Bertolotto}
\IEEEauthorblockA{\textit{School of Computer Science} \\ \textit{University College Dublin} \\
Dublin, Ireland \\
michela.bertolotto@ucd.ie}}

\maketitle

\begin{abstract}
The Information Retrieval research has used semantics to provide accurate search results, but the analysis of conceptual abstraction has mainly focused on information integration. 
We consider session-based query expansion in Geographical Information Retrieval, and investigate the impact of semantic granularity (i.e., specificity of concepts representation) on the suggestion of relevant types of information to search for. We study how different levels of detail in knowledge representation influence the capability of guiding the user in the exploration of a complex information space. 
A comparative analysis of the performance of a query expansion model, using three spatial ontologies defined at different semantic granularity levels, reveals that a fine-grained representation enhances recall. However, precision depends on how closely the ontologies match the way people conceptualize and verbally describe the geographic space.  
\end{abstract}


\begin{IEEEproof}
Copyright 2018 IEEE.  Personal use of this material is permitted.  Permission from IEEE must be obtained for all other uses, in any current or future media, including reprinting/republishing this material for advertising or promotional purposes, creating new collective works, for resale or redistribution to servers or lists, or reuse of any copyrighted component of this work in other works.
\end{IEEEproof}

\section{Introduction}
Several researchers have used semantic knowledge to support query expansion and reformulation in information search support; e.g.,  \cite{Wang-etal:17}. Moreover, in Ontology-Driven Geographic Information Systems, abstraction has been analyzed  to understand its ``potential for information retrieval at different levels of granularity'' \cite{Fonseca-etal:02b}. However, to the best of our knowledge, the impact of different domain conceptualizations on the accuracy of information search support has not been fully investigated yet. 
We focus on Geographical Information Retrieval, and aim at studying the influence of semantic granularity (i.e., the degree of specificity in concept representation) on query expansion. 
Specifically, we aim at measuring the extent to which, by combining general information about search behavior with a more generic, or a more detailed domain conceptualization, an automated system can learn regularities useful in the suggestion of concepts relevant to the user's information needs. 
For instance, by analyzing general search behavior we might discover that, if a user looks for kindergartens in a town, he or she might also be interested in information related to other children activities, such as play or sport areas. 

Notice that our aim is not only to measure the precision of the system's suggestions, but also to define a notion of ``richness" based on the number of relevant suggested concepts, as this is important for catalog exploration. 
Specifically, we investigate the following research questions: 

{\bf RQ1:}  \textit{What is the relationship between the semantic granularity of a domain conceptualization and the accuracy in suggesting types of information relevant to the user's needs during an exploratory search task?}

{\bf RQ2:} \textit{Does semantic granularity in the representation of geographical knowledge influence the richness of concept suggestion?}

In order to answer these questions, we need to know which concepts people frequently focus on during a search session, as a source of evidence of co-occurring information needs. Thus, we apply a session-based concept suggestion model to extract concepts co-occurrence, and we compare its performance by using different ontologies to train it and to interpret the search queries: a fine-grained ontology, a less detailed one developed for OpenStreetMap 
(www.openstreetmap.org) and the GeoNames Mappings ontology (www.geonames.org/ontology/mappings\_v3.01.rdf).
The results of an experiment based on a large query log reveal that a finer-grained semantic granularity improves recall and richness of results: with the first ontology the concept suggestion model achieves the best recall and supports the generation of more suggestions than with the other two. 
However, the model achieves the best precision and accuracy (F1) using the second ontology, which is based on crowdsourced data.

In the following, Section \ref{related} summarizes the related research.
Sections \ref{ontologies} and \ref{model} present the ontologies and the concept suggestion model we employed.
Section \ref{dataset} describes the datasets we used. 
Section \ref{experiments} describes our empirical evaluation.
Section \ref{conclusions} concludes the paper.

\begin{figure*}[h]
\centering
\captionsetup[subfigure]{position=b}
\begin{subfigure}[t]{0.5\textwidth}
  \centering
 \includegraphics[width=\linewidth]{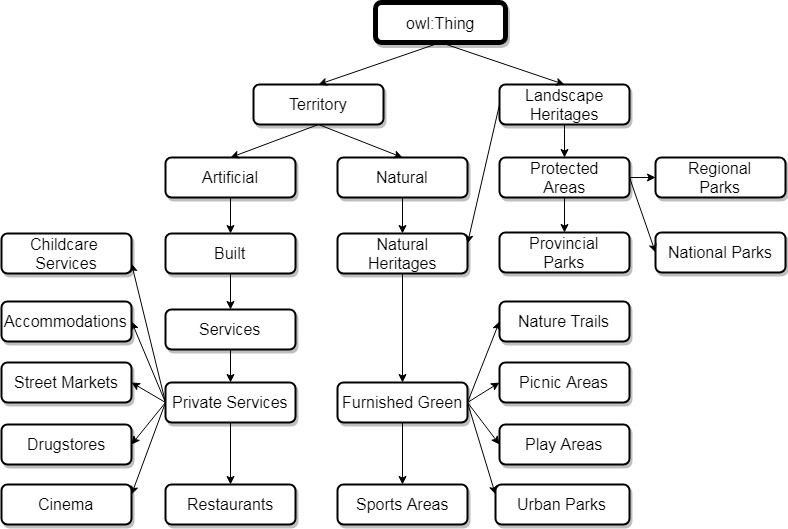}
  \caption{Portion of $N$ ontology.}\label{fig:N}
\end{subfigure}
\hfill
\begin{subfigure}[t]{0.48\textwidth}
  \centering
 \includegraphics[width=\linewidth]{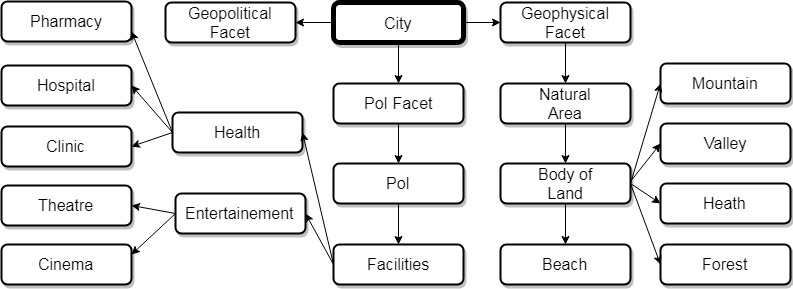}
  \caption{Portion of $L$ ontology.}\label{fig:L}
\end{subfigure} 
\par\bigskip 
\begin{subfigure}[t]{0.31\textwidth}
  \centering
 \includegraphics[width=\linewidth]{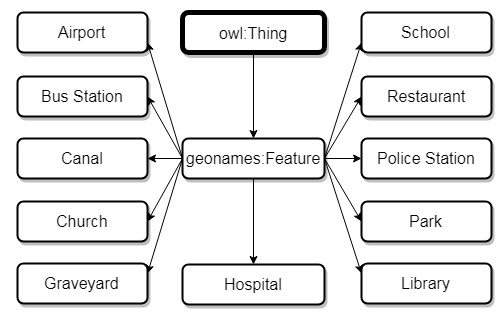}
  \caption{Portion of $GeoNames Mappings$ ontology. }\label{fig:Geonames}
\end{subfigure}
\caption{Ontologies selected for the empirical evaluation. The root nodes of the ontologies have thick borders for readability purposes.}
\end{figure*}

\section{Related Work}
\label{related}
Geographical Information Retrieval mainly focuses on the interpretation of the spatial component of a search task \cite{Palacio-etal:15,Henrich-Ludecke:07}.
However, Ballatore et al. \cite{Ballatore-etal:16} identify research themes and open questions, which include aspects related to search behavior models and semantic aspects of spatial search.
In this context, we focus on geospatial knowledge, and aim to identify the potential of semantic granularity in search support.

Term suggestion and query expansion have been based on the analysis of the lexical or semantic distance among terms \cite{Jones-etal:06,Fernandez-Reyes-etal:18}, on term co-occurrence in search logs \cite{Huang-etal:03,Chen-etal:08} and documents \cite{Gupta-Saini:17}, and on an analysis of click distribution on search results \cite{Cao-etal:08}. 
Jiang et al. propose a framework for search session derivation, geographic information extraction and geographic web search topic discovery, which exploits topic modeling techniques \cite{Jiang-etal:15}.
Differently, we aim at proposing concepts by using common patterns of concept exploration.

In Geographic Information Retrieval, ontologies have been used to improve geographic {\em features} extraction \cite{Laurini:15}.
%; e.g., see \cite{Laurini:15}. 
On the other hand, we employ them for geographic {\em concepts} extraction, in order to provide the user with topics to explore, rather than individual items. 

Researchers developed specific geographic ontologies, such as GeoNames. 
Moreover, they attempted to semantify geographical data targeted to specific tasks; e.g., the LinkedGeoData ontology links OpenStreetMap information to DBpedia, GeoNames, and others ontologies \cite{Janowicz-etalb:12}. 
Fonseca et al. propose an ontology to classify geographic elements with respect to geometrical features and attribute values; i.e., semantic features \cite{Fonseca-etal:02}. 
Finally, a relevant amount of research has been devoted to semantifying geospatial Open Data and crowdsourced data. For instance, OSMonto \cite{Codescu-etal:11} attempts to structure the shallow implicit ontology of OpenStreetMaps tags. 
Moreover, in \cite{Ballatore-etal:13} a Semantic Web resource extracted from the OpenStreetMap Wiki website is proposed, to semantify crowdsourced data. 

\begin{figure*}
 \centering
 \includegraphics[width=1.2\columnwidth]{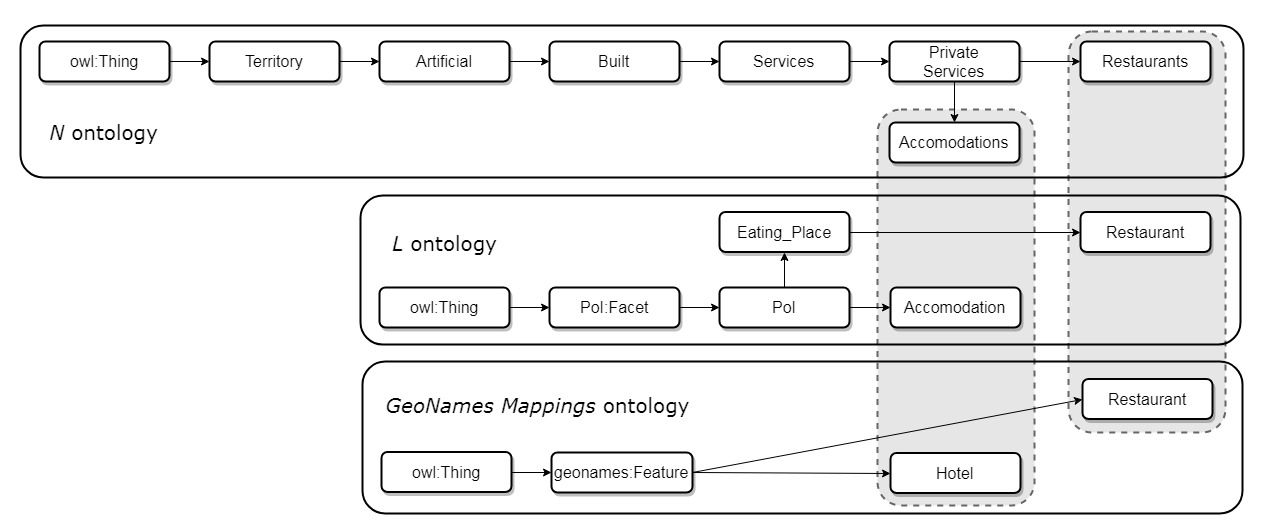}
 \caption{Portion of the three ontologies with an example of shared concepts (in gray areas).}
 \label{fig:euler}
\end{figure*}

\section{The Used Ontologies}
\label{ontologies}
In this section we analyze and compare the three geographical ontologies  used for the experiments. We focus on:
\begin{enumerate} 
\item 
Semantic granularity, i.e., the level of detail at which geographic objects are described \cite{Fonseca-etal:02b}. This is different from spatial granularity, which refers to the granularity of geographic toponyms; e.g., see \cite{Palacio-etal:15}.
\item 
Ontology structure, quantified in terms of number of concepts and number of subclass relations.
\end{enumerate}

\textbf{\textit{N} ontology} (see Figure \ref{fig:N}) contains fine-grained concepts about cities \cite{Voghera-etal:16}.
This ontology provides a multi-faceted specification of classes and relations that characterize the information space considering three high-level aspects: natural (e.g., parks), artificial (e.g., infrastructures) and normative (e.g., administrative boundaries). These aspects are specialized into classes at more than one level of detail. For instance, as partially reported in the figure, Landscape Heritage is specified into a deep hierarchy of classes in order to distinguish various types of Furnished Green from Protected Areas. The longest path between owl:Thing and the leaf classes has length = 6.

\begin{table}[t]
\centering
\caption{Structural comparison of the GeoNames, $L$ and $N$ ontologies.}
\begin{tabular}{lccc}
\hline
\textbf{Metric}  & \textbf{\begin{tabular}[c]{@{}c@{}}\textit{N}\end{tabular}} & \textbf{\begin{tabular}[c]{@{}c@{}}\textit{L}\end{tabular}} & \textbf{\begin{tabular}[c]{@{}c@{}}GeoNames\end{tabular}} \\ \hline
 \begin{tabular}[c]{@{}l@{}}Total number of  classes\end{tabular}      & 195  & 97   & 150                                                                   \\ %\hline 
 \begin{tabular}[c]{@{}l@{}}Total number of subclass relations\end{tabular} & 268  & 94   & 155 
\\ %\hline
\begin{tabular}[c]{@{}l@{}}Longest path to leaves (graph nodes)\end{tabular}
& 6  & 5   & 3 
\\ %\hline
\begin{tabular}[c]{@{}l@{}}Number of used classes\end{tabular}
 & 72   & 70   & 136                                                                  \\ %\hline
\begin{tabular}[c]{@{}l@{}}Number of used subclass relations\end{tabular}
  & 81   & 69   & 135                                                                  \\ %\hline
\begin{tabular}[c]{@{}l@{}}Longest path to used leaves (graph nodes)\end{tabular}
 & 6  & 5   & 2 
\\ %\hline
\begin{tabular}[c]{@{}l@{}}Mean degree of used classes (graph nodes)\end{tabular}
 & 4.5  & 3.63   & 67.5 \\  \hline
\end{tabular}
\label{tab:ontologiesMeasures}   
\end{table}

\textbf{\textit{L} ontology} (see Figure \ref{fig:L}) structures OpenStreetMap tags that refer to geographic concepts in the context of cities \cite{DiRocco:16}. It is multi-faceted and has a similar structure, but coarser granularity, than $N$; moreover, being derived from the largely-used OpenStreetMap repository, it reflects a conceptualization that comes from the general public. The geographic information about a city is described using three facets: Point of Interest (PoI), geoPolitical and geoPhysical. The longest path between the owl:Thing class and the leaf classes of \textit{L} has length = 5.

\textbf{\textit{GeoNames Mappings ontology}} (see Figure \ref{fig:Geonames}) maps the GeoNames taxonomy to external ontologies, such as schema.org and linkedgeodata.org.
GeoNames derives from the geonames.org database and describes more than 9,000,000 features categorized in 645 alphanumeric feature codes. The Mappings ontology maps  108 of its codes and describes knowledge at a coarse semantic granularity: information is organized in a flat structure including a top-level OWL class (owl:Thing), and a child class (geonames:Feature) that has all the GeoNames codes as its children. Only two subclasses of geonames:Feature, representing toponyms, have children themselves. 
Thus, even though the longest path between owl:Thing and the leaves of the ontology has length = 3, the path from owl:Thing to most classes has length = 2.

As a reference for the comparison among ontologies, we selected GeoNames Mappings (henceforth, GeoNames for brevity), which is largely adopted in the geospatial community. 
In each ontology, we only considered geospatial classes related to natural features of the territory (e.g, rivers, mountains) and artificial features (e.g., malls, buildings, streets). We excluded administrative boundaries, which concern spatial, rather than semantic, granularity and are abstract concepts.
The ontologies used in our analysis are thus subsets of the original ones.

Table \ref{tab:ontologiesMeasures} shows a set of metrics to describe the complete ontologies, and the subsets focused on natural and artificial features.
As shown in row ``Number of used classes", the used subset of GeoNames is the largest one, followed by the $N$ and the $L$ ones. 
The three ontologies share 8 concepts; GeoNames and $L$ share 23 ones; GeoNames and $N$ share 17 ones.

The subsets of $N$, $L$ and GeoNames have different semantic granularity. A rough measure of this can be obtained by jointly considering the length of the longest path between owl:Thing and the leaves of an ontology, and the mean degree of the nodes of the ontology graph. Among the three, $N$ has the maximum longest path (length = 6) and its mean degree is 4.5, which denotes the fact that several classes are specified into a relatively small set of subclasses. $L$ has few subclasses per node as well (mean degree = 3.63) but it defines concepts at a slightly more superficial level (length of longest path = 5). Finally, GeoNames is flat (length = 2) and has a very high mean node degree (67.5).
We report some examples to compare the ontologies. Class Park in GeoNames is represented in $N$ as  Regional\_Park,
\linebreak 
Provincial\_Park and National\_Park. Therefore, we can say that Park is shared between $N$ and GeoNames but they differ in granularity and thus the two representations can support concept identification at different levels of detail.
Similarly, Shopping of $L$  corresponds to Shopping\_mall, Market and Shop in $N$. 
Finally, in $N$, concept Restaurant is associated with a much longer and descriptive path than in GeoNames. The corresponding path in $L$ is descriptive but not as exhaustive as in $N$; see Figure~\ref{fig:euler}.

\section{Concept Suggestion Model (CS Model)}
\label{model}
In order to compare suggestion accuracy using different reference ontologies, we used the Concept Suggestion Model described in \cite{Mauro-Ardissono:18}. This model suits our needs because it extracts concept co-occurrence (w.r.t. term, or query co-occurrence, returned by other algorithms) from the search sessions of a query log, and thus enables us to identify information needs from a semantic point of view. 
The CS model bases concept suggestion on 
(i) the definition of co-occurrence clusters grouping concepts that are frequently referenced together in search sessions; 
(ii) the adoption of strategies that suggest concepts by matching the co-occurrence clusters to the search context according to different policies. By search context we mean the set of concepts referenced in the observed queries.

\subsection{Creation of Concept Co-occurrence Clusters}
\label{cluster-creation}
The process to create the concept co-occurrence clusters, starting from the query log and the ontology conceptualizing the domain knowledge, includes the following steps:
\begin{enumerate}
\item
Split the query log in sessions by aggregating queries according to their temporal proximity.
\item
Produce a {\em reduced dataset} by selecting the search sessions whose queries reference at least one concept corresponding to an ontology class.
For this task, the ontology classes are annotated with linguistic and domain knowledge. The lemmas of the words occurring in the queries are thus matched to the lemmas of the class annotations in order to find correspondences.
\item
Build a concepts co-occurrence graph by analyzing the sessions of the reduced dataset. 
In the graph, nodes denote concepts, edges denote concept co-occurrence in search sessions, and the weight of an edge denotes co-occurrence frequency of its input and output concepts. In order to focus on common search behavior, the edges having low weight are pruned. 
\item
Apply a community detection algorithm (COPRA \cite{Gregory:10}) to the graph to identify its concept communities. 
\end{enumerate}
The output of this phase is a set of clusters, each one consisting of an unordered set $CL = \{c_1, \dots, c_n\}$ of concepts $c_x$ that highly co-occur in the sessions of the reduced dataset.

\subsection{Concept Suggestion Strategies}
\label{strategies}
Given a search session $S = \{Q_1, \dots, Q_i, \dots, Q_n\}$, let $S@i$ be the observed portion of $S$, i.e., $\{Q_1, \dots, Q_i\}$. Moreover, let $C@i = \{c_1, \dots, c_k\}$ be the set of concepts identified by interpreting the queries of $S@i$. 

In \cite{Mauro-Ardissono:18}, three main strategies for cluster selection are proposed, which result in different concept suggestion policies:

{\bf SLACK:} 
select a set of clusters $\{CL_{x_1}, ..., CL_{x_y}\}$ that contain {\em at least} one concept $c$ such that $c \in C@i$; i.e., $c$ has been referenced in the observed portion of the session ($S@i$). 
For instance, given $C@i = \{c_1, c_3\}$, and three clusters, $CL_1 = \{c_1, c_3,c_7\}$, $CL_2 = \{c_2, c_3, c_5, c_8\}$, $CL_3 = \{c_5, c_8,c_9\}$,  SLACK selects $CL_1$ and $CL_2$, and it suggests $\{c_2, c_5, c_7, c_8\}$.

{\bf SLACK SELECTIVE:}
same as SLACK, but only one cluster that best matches $C@i$ is selected. We compute the degree of matching between a cluster $CL$ and $C@i$ as $|CL \cap C@i|$. In the example, the strategy suggests $\{c_7\}$ because the best matching cluster is $CL_1$.

{\bf STRICT:}
select the clusters $\{CL_{x_1}, \dots, CL_{x_y}\}$ containing {\em all} the concepts $c \in C@i$. In the example, the strategy does not suggest any concepts.

We aim at generating suggestions immediately after the first query of a search session; thus, we apply the strategies to $S@1$.

\section{Datasets for Training the CS Model}
\label{dataset}
We used the AOL query log\footnote{Retrieved from 
\url{https://archive.org/details/AOL\_search\_data\_leak\_2006}.} as a  source of search sessions. 
Although that log was involved in an information leak issue, we decided to use it for two reasons: firstly, our analysis is ethically correct, because we model general search behavior. The preference co-occurrence clusters we extract represent aggregate data about people behavior, and they abstract from the search histories of particular users. Secondly, to the best of our knowledge, the AOL log is the only public, large dataset that reports detailed data about textual search queries, and that can be used for linguistic interpretation.\footnote{We analyzed some public datasets but they did not meet our requirements. E.g., the Excite query dataset (\url{https://svn.apache.org/repos/asf/pig/trunk/tutorial/data/}) contains fewer queries (about 1M queries against the 20M of AOL log). In the Yahoo dataset (\url{https://webscope.sandbox.yahoo.com/catalog.php?datatype=l\&did=50}) 
queries are coded; thus, it is not possible to extract any linguistic information to learn the concept co-occurrence clusters.}

Each line of the AOL log represents a submitted query, or a click-through event, and reports an anonymous user id, the search query, date and hour when the query was submitted and other fields that are not relevant here.

\subsection{AOL-reduced Datasets}
To compare the impact of domain conceptualization on concept suggestion, we reduced the AOL query log using the $N$, $L$ and GeoNames ontologies. 
We generated three {\em AOL-reduced} datasets (AOL$_N$, AOL$_L$, AOL$_{GN}$) by filtering the sessions of the AOL log according to the concepts defined in each ontology; see Section \ref{cluster-creation}. For this task, we pre-processed the ontologies by automatically annotating their classes with linguistic knowledge (synonyms and linguistic definitions), using BabelFy multilingual Entity Linking and Word Sense Disambiguation service \cite{Moro-etal:14}, and Stanford CoreNLP lemmatizer \cite{Manning-etal:14}.

\begin{table}[t]
\centering
\caption{Size of \textit{AOL-reduced} datasets.}
\scalebox{1}{
%\scalebox{0.8}{
\begin{tabular}{lccc}
\hline
\textbf{Measure}            & \textbf{\begin{tabular}[c]{@{}c@{}}\textbf{AOL$_N$}\end{tabular}} & \textbf{\begin{tabular}[c]{@{}c@{}}\textbf{AOL$_L$}\end{tabular}} & \textbf{\begin{tabular}[c]{@{}c@{}}\textbf{AOL$_{GN}$}\end{tabular}} \\ \hline
Number of queries & 1,581,817  & 1,486,122 & 1,443,448                                                                   \\ 
Number of sessions & 945,945& 911,399 & 864,869\\ 
Number of users    & 247,868 & 240,474 & 232,931                                                               \\ \hline
\end{tabular}}
\label{tab:datasetsCharacteristics}  
\end{table}

As shown in Table \ref{tab:datasetsCharacteristics}, the AOL$_N$ dataset contains the highest number of queries, sessions and users. AOL$_L$ contains the second highest number of queries, sessions and users, and AOL$_{GN}$ is the smallest dataset, even though the GeoNames ontology is larger than the other two. 
We explain this finding using semantic granularity: $N$ is the most detailed ontology and supports the selection of the largest number of queries from the log. However, in GeoNames, the coarse-grained classes limit the identification of concepts in search queries, and make the selection of relevant sessions more constrained than using $L$.
The datasets have some similarities. As reported in Table \ref{tab:sessionLength}, the mean length of their sessions is almost the same: 1.67 queries in AOL$_N$ and AOL$_{GN}$, 1.63 in AOL$_L$. Moreover, the distribution of sessions w.r.t. their length follows a Power Law (most sessions are short, few contain many queries). Thus, we can use them to compare the concept suggestion strategies.    

\begin{table}[t]
\centering
\caption{Length of the search sessions of the \textit{AOL-reduced} datasets.}
\scalebox{1}{\begin{tabular}{lccc}
\hline
\textbf{Measure}                                                                         & \textbf{\begin{tabular}[c]{@{}c@{}}\textbf{AOL$_{N}$}\end{tabular}} & \textbf{\begin{tabular}[c]{@{}c@{}}\textbf{AOL$_{L}$}\end{tabular}} & \textbf{\begin{tabular}[c]{@{}c@{}}\textbf{AOL$_{GN}$}\end{tabular}} \\ \hline   
\begin{tabular}[c]{@{}l@{}}Min number of queries per session\end{tabular}   & 1  & 1 & 1\\ %\hline
\begin{tabular}[c]{@{}l@{}}Max number of queries per session\end{tabular}   & 93& 96 & 103\\ %\hline
\begin{tabular}[c]{@{}l@{}}Mean number of queries per session\end{tabular}     & 1.67  & 1.63 & 1.67\\ %\hline
\begin{tabular}[c]{@{}l@{}}Median number of queries per session\end{tabular} & 1  & 1 &1\\ %\hline
Standard deviation  & 1.45  & 1.38 & 1.44\\ \hline
\end{tabular}}
\label{tab:sessionLength} 
\end{table}

\begin{table}[t]
\centering
\caption{Sample concept co-occurrence clusters.}
\begin{tabular}{lc}
\hline
\textbf{Dataset} & \textbf{Sample clusters}   \\ \hline
\textbf{AOL$_N$} & {\begin{tabular}[c]{@{}c@{}}
	\{Play Areas, Sport Areas, Kindergartens,\\ Law Enforcement, Hospitals\} \\
	\{Play Areas, Sport Areas, Libraries,\\ Childcare services, Law Enforcement, Schools\}\end{tabular}}     \\ \hline
\textbf{AOL$_L$} & {\begin{tabular}[c]{@{}c@{}}
	\{Educational\_institution, Kindergarten,  University\} \\
	\{Bar, Cafe, Eating\_place, Pub, Restaurant\}\end{tabular}}    \\ \hline
\textbf{AOL$_{GN}$}    & {\begin{tabular}[c]{@{}c@{}}	
	\{Library, Museum, Police\_Station, University\} \\
    \{Beach, Resort\}
 \end{tabular}}     \\ \hline
\end{tabular}
\label{t:clusters}
\end{table}

\subsection{Concept Co-occurrence Clusters}
\label{clusters}
For each ontology $X$, and AOL-reduced dataset {\em AOL$_X$}, we created a concept co-occurrence graph $G_X$ as described in Section \ref{cluster-creation}. Also in this case, we pruned $G_X$ to remove the weakest co-occurrence relations among concepts. 
%Specifically,  we selected a threshold for pruning the arcs of $G_X$ to optimize the prediction accuracy of the resulting concept co-occurrence clusters. 
Then, we applied the COPRA algorithm to $G_X$ to identify the communities of concepts that frequently co-occur in the search sessions. The algorithm returned 23 clusters for AOL$_N$, 19 for AOL$_L$ and 23 for AOL$_{GN}$. Table \ref{t:clusters} shows two sample clusters for each dataset.

The clusters generated from AOL$_N$ contain a larger number of concepts than the others (about 21\% more than those produced using AOL$_L$, and about 31\% more than those derived from AOL$_{GN}$). This finding is in line with the granularity of the ontologies: being more specific, $N$ makes it possible to identify a more diverse set of co-occurring concepts in the search sessions. In contrast, GeoNames supports the identification of few concepts, and thus produces small clusters. 

\begin{table*}[t]
\centering
\begin{subtable}[t]{1\columnwidth}
\centering
\caption{SLACK performance ($S@1$).}
%\scalebox{1}{
\begin{tabular}{lccc}
\hline
\textbf{}                                                                    & \textbf{\begin{tabular}[c]{@{}c@{}}{\em AOL}$_N$ \end{tabular}} & \textbf{\begin{tabular}[c]{@{}c@{}}{\em AOL}$_L$\end{tabular}} & \textbf{\begin{tabular}[c]{@{}c@{}}{\em AOL}$_{GN}$\end{tabular}} \\ \hline
\begin{tabular}[c]{@{}l@{}}Min number of  suggested concepts\end{tabular}  & 0  & 0  & 0                                                                    \\ %\hline
\begin{tabular}[c]{@{}l@{}}Max number of  suggested concepts\end{tabular}  & \textbf{7}  & 4 & 4                                                                    \\ %\hline
\begin{tabular}[c]{@{}l@{}}Mean number of  suggested concepts\end{tabular} & \textbf{2.688} & 0.496 & 1.371                                                                 \\ %\hline
Recall                                                                      & \textbf{0.791}  & 0.783  & 0.768 
\\ %\hline
Precision                                                                    & 0.614  & \textbf{0.882} & 0.712                                                                 \\ %\hline
F1                                                                           & 0.691  & \textbf{0.829} & 0.739                                                                 \\ \hline
\end{tabular}
%%}
\label{SLACK}
\end{subtable}
\begin{subtable}[t]{1\columnwidth}
\centering
\caption{SLACK-SELECTIVE performance ($S@1$).}
%\scalebox{1}{
\begin{tabular}{lccc}
\hline 
\textbf{}                                                                    & \textbf{\begin{tabular}[c]{@{}c@{}}{\em AOL}$_N$ \end{tabular}} & \textbf{\begin{tabular}[c]{@{}c@{}}{\em AOL}$_L$\end{tabular}} & \textbf{\begin{tabular}[c]{@{}c@{}}{\em AOL}$_{GN}$\end{tabular}} \\ \hline 
\begin{tabular}[c]{@{}l@{}}Min number of  suggested concepts\end{tabular}  & 0  & 0  & 0                                                                    \\ %\hline
\begin{tabular}[c]{@{}l@{}}Max number of  suggested concepts\end{tabular}  & \textbf{5}  & 4 & 3                                                                    \\ %\hline
\begin{tabular}[c]{@{}l@{}}Mean number of suggested concepts\end{tabular} & \textbf{2.309} & 0.455 & 1.054                                                                 \\ %\hline
Recall                                                                       & \textbf{0.783}  & 0.759  & 0.761 
\\ %\hline
Precision                                                                    & 0.621  & \textbf{0.875} & 0.725                                                                 \\ %\hline
F1                                                                           & 0.692  & \textbf{0.813} & 0.743                                                                 \\ \hline
\end{tabular}
% }
\label{SLACK-SELECTIVE}
\end{subtable}
\caption{Performance of the concept suggestion strategies applied to the first query ($S@1$) of the sessions of the {\em AOL-reduced} datasets.}
\label{tab:results}
\end{table*}

The clusters generated using AOL$_L$ are more topic-centered than the other ones: the concepts of each AOL$_L$ cluster have the same parent concept in the $L$ ontology.
As $L$ is fairly fine-grained, it describes different topics as non-leaf concepts of the ontology. Thus, having the same parent means being strictly related from a semantic point of view.
Indeed, AOL$_N$ generates a few topic-centered clusters, as well. However, most of its clusters include semantically distant concepts that typically descend from different high-level concepts of the ontology. For instance, in Table \ref{t:clusters}, Play Areas and Sport Areas descend from Natural, a high-level class of the $N$ ontology. The other concepts of the cluster descend from Artificial (another high-level class). 
Finally, the clusters generated from AOL$_{GN}$ cover a broad range of topics, given the flat structure of the GeoNames ontology.

\section{Empirical Evaluation}
\label{experiments}
We measured the performance of the CS Model on the AOL-reduced datasets, using the respective concept co-occurrence clusters for concept suggestion. For each dataset, we measured the richness and the accuracy of the suggestions that the SLACK, SLACK-SELECTIVE and STRICT strategies generated, by matching the concepts they proposed, given the first query of a session $S$ ($S@1$), against those referenced in the remainder of $S$; i.e., those actually explored by users.
For the evaluation we applied 10-fold cross-validation on each dataset, after having randomly distributed its sessions on folders. We used 90\% of the sessions as learning set and 10\% as test set. 
 
Table \ref{tab:results} presents the evaluation results obtained by applying SLACK and SLACK-SELECTIVE to the {\em AOL-reduced} datasets, considering only the sessions that include at least two queries. We omit the results of STRICT due to space limitations.

Henceforth, $C_S$ denotes the set of concepts explored by the user in a session $S$; $C_{Sugg_{S@i}}$ is the set of concepts suggested by a strategy after analyzing the $S@i$ portion ($i = 1$).

{\bf Richness of suggestions}. The first 3 lines of Tables \ref{SLACK} and \ref{SLACK-SELECTIVE} compare the numbers of relevant concepts suggested by the strategies, per dataset. These numbers measure the suggested concepts that are explored by the users, and thus represent useful suggestions. It can be seen that SLACK recommends more relevant concepts than SLACK-SELECTIVE. However, both strategies obtain the highest values when they are applied to AOL$_N$, which is the most detailed ontology.

{\bf Recall ($R@i$).} This measure represents the percentage of concepts explored in search sessions that are suggested by the strategies. 
It represents the system capability to cover the topics that users actually explore in their searches. Tables \ref{SLACK} and \ref{SLACK-SELECTIVE} show that strategies achieve the best $R@1$ with AOL$_N$.

{\bf Precision ($P@i$)} represents the percentage of suggested concepts that are explored in the remainder of the sessions.
Tables \ref{SLACK} and \ref{SLACK-SELECTIVE} show that the strategies achieve the best $P@1$ with AOL$_L$, and this value is much higher than those obtained on the other datasets.

\begin{figure}[t]
  \centering
  \begin{subfigure}[t]{0.95\columnwidth}
  \centering
 \includegraphics[width=0.95\columnwidth]{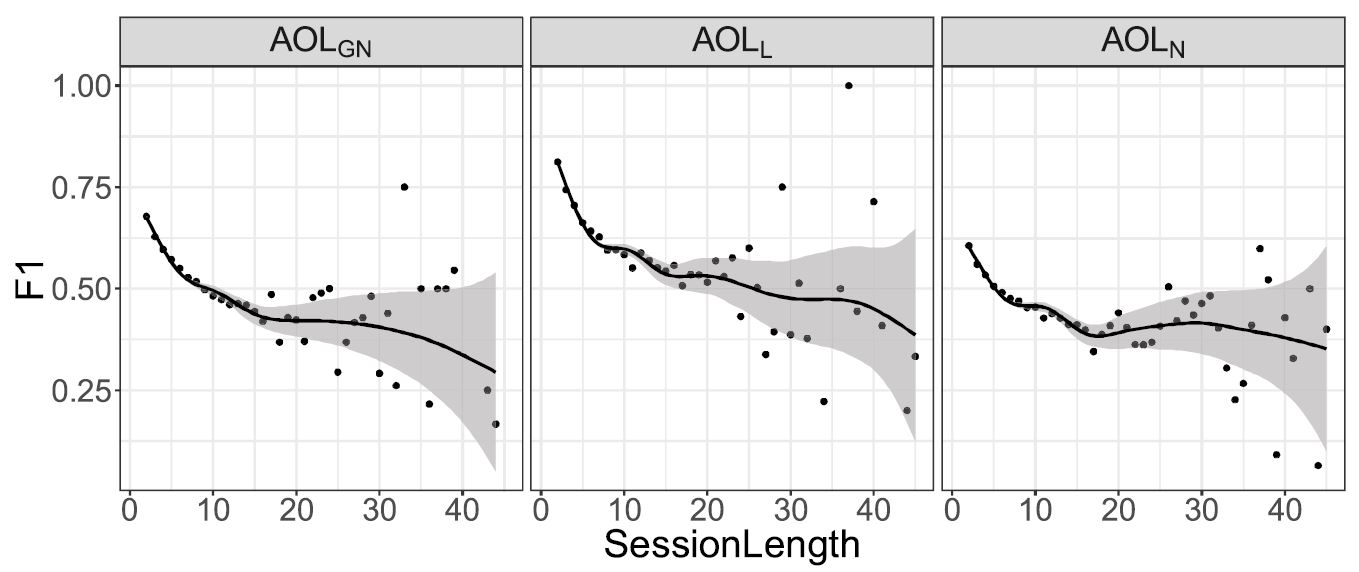}
 \caption{SLACK}
 \end{subfigure}
 \hfill

\begin{subfigure}[t]{0.95\columnwidth}
  \centering
   \includegraphics[width=0.95\columnwidth]{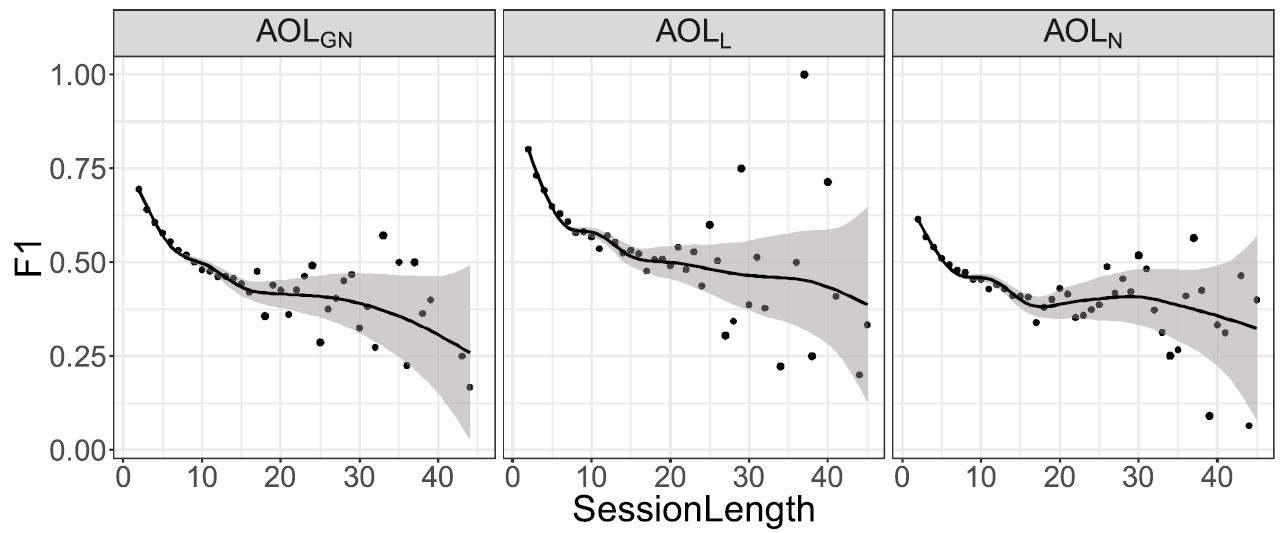}
   \caption{SLACK-SELECTIVE}
   \end{subfigure}
  \caption{Distribution of F1 score with respect to the length of sessions.
  }\label{fig:session_mean}
\end{figure}

{\bf Accuracy ($F1@i$)} integrates precision and recall.
Both strategies obtain the highest $F1@1$ score on AOL$_L$. 
Figure \ref{fig:session_mean} shows the variation of $F1@1$ w.r.t. the length of search sessions (length $>=$ 2). The points denote the mean value of F1 for each length value. The black lines represent the local regression on this data  and show the trend of F1: at first, it decreases; then the regression line stabilizes. For longer sessions, the values of F1 do not correspond to any trend. Thus, the confidence of the regression line (in gray) decreases. 

The results of the evaluation contribute to answer our research questions. 
Regarding {\bf RQ1}, results support the hypothesis of a dependency between the semantic granularity of a domain conceptualization and the accuracy in the suggestion of concepts relevant to a user search. 
Specifically, a finer-grained semantic granularity enhances recall by supporting the generation of larger concept co-occurrence clusters that have a broad scope (they can refer to semantically distant concepts of the ontology). Thus, given a set of concepts identified in the observed query, the selection of one or more clusters that match it provides a larger pool of concepts to be suggested. 

Precision increases when using an ontology that covers a broad range of concepts, like GeoNames. However, it depends on how close the ontology is to the way people conceptualize and refer to geographic information. The evaluation shows that the suggestions obtained with the crowdsourced $L$ ontology, which is more specific than GeoNames, but less than $N$, are more precise than those obtained with the other ontologies.

Regarding {\bf RQ2}, the results support the hypothesis that the semantic granularity of a domain conceptualization impacts on the richness of concept suggestions.
A finer-grained ontology supports the suggestion of a larger set of relevant concepts because it is able to match a larger variety of concepts in the query logs. Moreover, it is associated with the largest clusters; thus, given a search context, it has higher suggestion capability.

\section{Conclusions}
\label{conclusions}
We investigated how semantic granularity in geographical knowledge conceptualization influences concept suggestions in information search support.
We found that a finer-grained domain conceptualization supports the suggestion of a larger set of concepts, with higher recall. However, the relationship between the formal conceptualization (ontology) and the way people conceptualize and verbally describe geographic space influences precision and, consequently, overall accuracy.

\bibliographystyle{IEEEtran}
% Generated by IEEEtran.bst, version: 1.12 (2007/01/11)

\end{document}